# Significant Dzyaloshinskii-Moriya Interaction at Graphene-Ferromagnet Interfaces due to Rashba-effect

Hongxin Yang[1,2*,#], Gong Chen[3*,#], Alexandre A.C. Cotta[3,4,5], Alpha T. N'Diaye[3], Sergey A. Nikolaev[1], Edmar A. Soares[5], Waldemar A. A. Macedo[4], Andreas K. Schmid[3], Albert Fert[2] & Mairbek Chshiev[1]

[1] Univ. Grenoble Alpes, INAC-SPINTEC, 38000 Grenoble, France; CNRS, SPINTEC, 38000 Grenoble, France; CEA, INAC-SPINTEC, 38000 Grenoble, France

[2] Unité Mixte de Physique, CNRS, Thales, Univ. Paris-Sud, Université Paris-Saclay, Palaiseau 91767, France

[3] NCEM, Molecular Foundry, Lawrence Berkeley National Laboratory, Berkeley, California 94720, USA

[4] Centro do Desenvolvimento da Tecnologia Nuclear, CDTN, 31270-901 Belo Horizonte, MG, Brazil

[5] Departamento de Física, ICEx, Universidade Federal de Minas Gerais, 31270-901 Belo Horizonte, MG, Brazil

\* These authors contributed equally to the work.

[#] To whom correspondence should be addressed.

E-mail: hongxin.yang.spintec@gmail.com, gchenncem@gmail.com

**The possibility of utilizing the rich spin-dependent properties of graphene has attracted great attention in pursuit of spintronics advances. The promise of high-speed and low-energy consumption devices motivates a search for layered structures that stabilize chiral spin textures such as topologically protected skyrmions. Here we demonstrate that chiral spin textures are induced at graphene/ferromagnetic metal interfaces. This is unexpected because graphene is a weak spin-orbit coupling material and is generally not expected to induce sufficient Dzyaloshinskii-Moriya interaction to affect magnetic chirality. We demonstrate that graphene induces a new type of Dzyaloshinskii-Moriya interaction due to a Rashba effect. First-principles calculations and experiments using spin-polarized electron microscopy show that this graphene-induced Dzyaloshinskii-Moriya interaction can have similar magnitude as at interfaces with heavy metals. This work paves a new path towards two-dimensional material based spin orbitronics.**

The unique properties of graphene including well-defined single atomic layer thickness, massless linear dispersion of its electronic structure, and long spin diffusion length have motivated the search for graphene-based phenomena that may enable spintronic applications[1,2,3]. Recently, graphene was shown to play key roles in several magnetic phenomena, including graphene-based tunnel magnetoresistance[4,5,6], enhancement of the spin-injection efficiency[7,8], Rashba effect[9,10], quantum spin Hall effect[11] and large perpendicular magnetic anisotropy (PMA)[12].

At the same time, recent progress in the field of spin orbitronics was stimulated by discoveries of phenomena permitting highly efficient electrical control of chiral spin textures, e.g. fast domain wall (DW) dynamics[13,14,15,16] and skyrmion motion at ultralow current densities[17,18,19,20]. These findings hold promise for applications in memory[21,22,23] and logic devices[24] where the interfacial Dzyaloshinskii-Moriya Interaction (DMI)[25,26] has been recognized as a key ingredient in creation, stabilization, and manipulation of skyrmions[27,28,29,30,31,32] and chiral DWs[33,34]. While chiral magnetism induced by the interfacial DMI has become an important topic, the DMI at interfaces with graphene was not expected to be significant because, according to the Fert-Levy model[35], the DMI scales with spin-orbit coupling (SOC) in the material contacting the ferromagnetic metal (FM) layer[36] and graphene lacks strong SOC. Recent results reported the observation of enhanced PMA at the graphene/Co interface, even though strong interfacial PMA is also often associated with strong SOC. This suggests that graphene/FM interfaces are unusual: if graphene enhances the PMA at interfaces in the absence of strong SOC, then it is interesting to ask if graphene has similarly strong effects on the DMI helping thereby to promote this and other 2D materials for spin orbitronics. In the following, this idea is tested by exploring the interfaces of graphene with cobalt and nickel, where these two FM elements are chosen for the small lattice mismatch and strong interaction with graphene.

**First-principles calculations**

The structures of graphene/FM films modelled here are shown in Fig. 1, where a layer of graphene coats the surfaces of three-monolayer (ML) thick hcp Co(0001) and fcc Ni(111) films. Arrows schematically indicate clockwise/right-handed and anticlockwise/left-handed (in parenthesis) spin spiral chirality. The calculated ground state structure is consistent with previous reports[4,12], where one carbon atom of the graphene unit cell is located on top of the adjacent Co(Ni) atom and another carbon atom is located above the hollow site, with the graphene/Co(Ni) distance of about 2.12 (2.15) Å.

We use the chirality dependent total energy difference approach applied previously for Co/Pt structures[31,36,37] to calculate microscopic and micromagnetic DMI constants, $d^{tot}$ and $D$, respectively, as well as the layer-resolved DMI, $d^k$, where $k$ indicates the individual atomic layers within FM films. As one can see from Fig. 2 for the calculated results, the largest DMI can reach up to 1.14 meV per atom for a graphene coated single atomic layer of Co, while for 2 and 3 ML of Co films coated by graphene, the amplitude of $d^{tot}$ drops to 0.16 and 0.49 meV, respectively (Fig. 2a). Moreover, $d^{tot}$ of graphene/Co (brown bars in Fig.2a) is generally stronger than that of graphene/Ni (green bars in Fig. 2a) for all thicknesses considered. For the micromagnetic DMI, $D$, we found that its magnitude decreases as a function of the FM layer thickness for both graphene coated Co and Ni films, due to interfacial origin of the DMI leading to the inverse proportionality with respect to FM layer thickness.[36]

In order to elucidate the origin of such a significant DMI in graphene coated FM, we then calculated the layer-resolved DMI, $d^k$, and associated SOC energy difference, $\Delta E_{SOC}^k$, for the case of graphene coated 3ML Co films. Fig. 2c shows that the largest layer-resolved DMI, $d^k$, is located at the interfacial Co layer, labelled as Co1 (blue bar), which is in contact with graphene, while within the layers further from the interface the DMI decays very fast (red and black bars) similarly to previously reported case at Co/Pt interface[36]. However, significant differences between graphene/Co and Co/Pt emerge in terms of where the corresponding SOC energy source is located. As shown in Fig. 2d, the largest associated SOC energy difference, $\Delta E_{SOC}^k$, originates from the same Co1 layer rather than from the non-magnetic side of the interface, where it is almost zero. This is drastically different from the Co/Pt case where the SOC energy difference is mainly contributed by the adjacent Pt layer. These findings indicate that the physical mechanism governing the strength of the DMI in graphene/Co interface is very different from that in Co/Pt, which is captured by the Fert-Levy model[35,36]. Instead, in graphene/Co the dominating mechanism is the Rashba-type DMI. According to the latter[38,39,40], the DMI parameter can be roughly expressed as $d = 2k_R A$, where $A$ is the exchange stiffness and $k_R = \frac{2\alpha_R m_e}{\hbar^2}$ is determined by the Rashba coefficient, $\alpha_R$, and effective electron mass, $m_e$. The latter in Co was measured to be about 0.45 $m_0$[41] (with $m_0$ being the rest mass of electron), and the exchange stiffness, $A$, was found to be about 15.5 ×10$^{-12}$ J/m[32,42]. The Rashba coefficient, $\alpha_R$, can then be extracted from $\alpha_R=2E_0/k_0$, where $E_0$ is the Rashba splitting at the wave vector $k_0$. We calculated the Rashba splitting for graphene/Co(3ML) slab by switching on SOC and putting the magnetization along <11$\bar{2}$0> and <$\bar{1}\bar{1}$20>. As one can see in Figs. 2e and f, the corresponding band shifts are a signature of the Rashba effect even though it deviates slightly from the conventional linear dependence given by $\alpha_R$ (**σ** x **k**)·**z**. Different characters of the band splitting at the $\bar{\Gamma}$ point can be attributed to the fact that Co $d$ orbitals are influenced by different potential gradients due to the polarization between graphene and Co that provides an intrinsic electric field and considerably enhances the effective value of SOC at the interface. We chose a band close to the Fermi level at $\bar{\Gamma}$ point, as shown in Fig. 2f, to estimate the Rashba-type DMI. The Rashba splitting, $E_0$, is about 1.28 meV at $k_0$=0.031 Å$^{-1}$, and the Rashba coefficient, $\alpha_R$ is thus found to be about 82.6 meV· Å. This leads to $k_R$=9.8×10$^{-3}$ Å$^{-1}$ and therefore $d$=0.30 meV, which is a bit smaller than the value calculated from first-principles, $d$=0.49 meV for graphene coated 3 ML Co films. The reason for the smaller DMI value extracted from the Rashba effect can be ascribed to the fact that the Rashba-type DMI was estimated by using only one band close to Fermi level. As reported in recent studies[43], the magnitude and sign of $\alpha_R$ is generally band-dependent due to band-specific orbital orderings of the orbital angular momentum giving rise to the band-dependent orbital chirality.

**Experimental observation of graphene-induced DMI**

Experimental tests of the DMI were done by using spin-polarized low-energy electron microscopy (SPLEEM), by direct imaging DWs in perpendicularly magnetized films (see

Methods). The sign of the DMI can be determined by observing the chirality of DWs[30,34,44], while the strength of the DMI vector, *d*, can be quantified by measuring the film thickness dependence of a transition from chiral Néel walls (in thin films, where the interfacial DMI influences DW texture) to achiral Bloch walls (in thicker films, where dipolar forces outweigh the DMI)[34,44]. We cannot prepare a free standing graphene/Co bilayer where the thickness of Co is several ML, instead, high quality Co/graphene samples were prepared on top of Ru(0001) single-crystal substrates (See Methods).

Figs. 3a,b show compound SPLEEM images highlighting the DW spin structure in graphene/Co/Ru(0001) films, where black and grey shades indicate that the magnetization is perpendicular to the film plane with $+M_z$ and $-M_z$ vectors, respectively, while colours represent the in-plane magnetization vector according to the colour wheel (inset). For Co thickness of 3.9 ML (Fig. 3a) the in-plane component of the magnetization within DWs (white arrows) is perpendicular to the DW tangent, and always points from black domains to grey domains, i.e. from $+M_z$ and $-M_z$: this indicates that the DWs have a left-handed/anti-clockwise chiral Néel texture[34,44]. For Co thickness of 8.4 ML (Fig. 3b), the magnetization vector within DWs is aligned parallel to the DW tangent: this indicates that the DW has a Bloch-type texture. Moreover, the magnetization vector within these DWs reverses its direction in a number of places, indicating that these DWs are achiral Bloch-walls[45]. This thickness-dependent transition of the DW type and chirality can be tracked in more detail using histogram as plotted in Fig. 3c (see Methods). The histogram represents the distribution of the angle *α*, defined as the angle between the DW magnetization vector **m** and the normal direction of DW, **n** (Fig. 3c inset). The distribution of the angle α gradually evolves from a single peak around 180º for Co 3.9 ML to double peaks at ±90° for Co 8.4 ML thicknesses.

The strength of the DMI in this system can be estimated as *d*=0.11±0.04 meV per atom (Fig. 3g), by computing the dipolar energy penalty of Néel-textured DWs at the film thickness threshold of the transition to Bloch-textured DWs (see details in Refs. 34 and 44). Note that the value of *d* contains contributions from both, the graphene/Co interface and Co/Ru interface, and the DMI at Co/Ru needs to be tested so that the DMI at graphene/Co can be deduced. Due to complicated strain-related spin-reorientation transitions in ultrathin Co films on Ru(0001)[46], we measured the Co/Ru DMI by growing [Ni/Co]$_n$ multilayers on Ru substrates. SPLEEM images (Fig. 3d,e) and the corresponding *α* histograms (Fig. 3f) show that the [Ni/Co]$_n$/Ru system contains right-handed/clockwise chiral Néel walls for *n*=2 and achiral Bloch walls for *n*=4: from these observations the DMI at Co/Ru is quantified[34,44] as *d*=-0.11±0.06 meV per atom (Fig. 3g), which has opposite sign compared to the DMI of graphene/Co/Ru trilayers. This experimental value is consistent with our first-principles calculations, *d*=-0.17 meV per atom for Co[3ML]/Ru[3ML]. Since the DMI is very localized at the interface[36,37], one can deduce that the DMI at the graphene/Co interface is opposite and is about twice as strong as the DMI at the Co/Ru interface. More quantitatively, our experiments suggest that the DMI for the graphene/Co interface has a value of about *d*=0.22±0.1 meV per atom (see Fig. 3g), consistent with the calculated DMI of *d*=0.49 meV for graphene/Co[3ML], as shown in Figs. 2 and 3.

**Tailoring a giant DMI in graphene-based heterostructures**

It was previously proposed that the DMI can be amplified using multilayer structures[32,34,37,47]. As summarized in Fig. 2, the sign of the DMI for graphene/Ni with Ni thickness of 1 and 2 MLs is negative (clockwise/right-handed chirality), while for graphene/Co the sign is always positive (anticlockwise/left-handed chirality). This suggests the possibility to obtain large DMI values by building ternary superlattices based on graphene/Co/Ni heterostructures. We tested this hypothesis with first-principles calculations by modelling graphene/[Co/Ni/graphene]$_n$ structures (Fig. 4). The calculated value of *d* increases with respect to the number of repeating units, *n*, with a slope less than one. Further calculations indicate that the amplification of the DMI can be further enhanced in Van der Waals heterostructures where two FM layers are separated by two MLs of graphene, i.e. in multilayers of the graphene/[Co/Ni/bilayer-graphene/]$_{(m-1)}$/Co/Ni/graphene structure. The result obtained for *m*=2 with *d*=1.13 meV suggests that in multilayers of *n* repeating units the DMI approaches a value of *m* times the DMI of a single graphene/Co/Ni/graphene unit. Furthermore, calculating the PMA for graphene/[Co/Ni/graphene]$_n$ heterostructures shows a linear increase with the number of repeating units *n*, that is similar to the behaviour of graphene/[Co/graphene]$_n$ reported before[12].

From the values of the DMI at Co/graphene interfaces obtained in this work, we predict that graphene induced DMI should be sufficient to stabilize magnetic chiral spin textures in ultrathin FM films attached to graphene. For instance, magnetic chiral DWs and skyrmions have been observed in weak DMI systems (-0.12 meV per atom at Ni/Ir interface[44], or 0.15 meV per atom in Fe/Ni/Cu system[30,34]). The proposed [Co/Ni/graphene]$_n$ heterostructure allows simultaneous enhancement of the DMI and PMA, which may be helpful for stabilizing chiral spin textures such as skyrmions with an extremely small size. Moreover, graphene/Co(Ni) grown on copper could be interesting since graphene production on copper is a well-established process[48], where the graphene related interface is expected to dominate the DMI due to the ignorable DMI at Co(Ni)/Cu interface[47].

In summary, we have discovered both from first-principles calculations and from magnetic imaging experiments that graphene/FM interface generates significant DMI. We showed that the physical origin of this DMI is the Rashba-effect. The discovery of the DMI induced by graphene along with its distinctive electronic properties[49], enhancement of PMA[12], and its ability to act as an excellent capping layer[50], may open up a new area in the field of spintronics.

**METHODS**

**First-principles calculations.** The Vienna *ab initio* simulation package (VASP) was used in our calculations with electron-core interactions described by the projector augmented wave method, and the exchange correlation energy calculated within the generalized gradient approximation of the Perdew-Burke-Ernzerhof (PBE) form[51,52]. The cutoff energies for the plane wave basis set used to expand the Kohn-Sham orbitals were chosen to be 520 eV for all calculations. The Monkhorst-Pack scheme was used for the Γ-centred 4×16×1 *k*-point mesh. In order to extract the DMI vectors, the calculations were performed in three steps. First, the corresponding structures were relaxed until the forces become smaller than 0.001 eV/Å to determine the most stable interfacial geometries. In our DMI calculations, we coated 1 to 3 monolayers of hcp Co(0001) or fcc Ni(001) films by graphene in a 4 by 1 surface unit cell with π/2 spin rotations (Fig. 1). Next, the Kohn-Sham equations were solved with no spin-orbit interaction taken into account to find out the charge distribution of the system's ground state. Finally, spin-orbit coupling was included and the self-consistent total energy of the systems was determined as a function of the constrained magnetic moments. We employ the same method used for DMI calculations in frustrated bulk systems and insulating chiral-lattice magnets[53] and adapted to the case of interfaces. As for the Rashba effect, we adopted the same approach as in Ref. [54] (see also Supplementary Fig. S.1 and corresponding discussion).

**Sample preparation.** We conducted the experiments in the SPLEEM system at National Center for Electron Microscopy of Lawrence Berkeley National Laboratory. All samples were prepared under ultra-high vacuum (UHV) conditions, with base pressure better than $4.0 \times 10^{-11}$ Torr. Ru(0001) substrates were cleaned by repeated flash annealing at 1470 K in $3.0 \times 10^{-8}$ Torr $O^2$ atmosphere and final annealing at 1430 K under UHV. After such procedure, we did not observe any trace of contaminants by Auger electron spectroscopy (AES) and LEEM. Furthermore, high-quality low energy electron diffraction patterns were obtained, indicating a well-ordered surface.

Graphene was grown by chemical vapour deposition method, where we kept the substrate at 920 K under ethylene atmosphere ($10^{-8}$ Torr) for around 15 minutes, while observing the process by LEEM. Preparing graphene at low growth temperature is required for a good intercalation process, since defects within the graphene layer assist the cobalt migration. The presence of graphene was confirmed by the moiré pattern in low energy electron diffraction[55] (see Supplementary Fig. S.2). After cooling graphene/Ru(0001) to room temperature, an amount of one monolayer Co was deposited by electron beam evaporation at rates of 0.18 ML per minute, and intercalated by annealing at 620 K for 3 minutes[56]. In order to achieve higher Co thicknesses, we repeated the intercalation of additional monolayer-doses of Co, exploring layer thicknesses up to 24 ML Co. The Co growth rate was calibrated by monitoring the LEEM image intensity during the deposition of Co directly onto bare Ru (0001).

Ni and Co layers on Ru (0001) were prepared in a similar way, where film thicknesses were calibrated by following the image intensity oscillations associated with atomic layer-by-layer growth. After the deposition process, the [Ni/Co]$_n$ multilayers were annealed at the same temperature as the Co intercalation through graphene, to directly compare two cases. The annealing of [Ni/Co]$_n$/Ru(0001) multilayers does not influence the magnetic chirality, as shown in Supplementary Fig. S.3.

Possible sign of Co diffusion into Ru was monitored by X ray photoelectron spectroscopy (XPS) in the Co/Ru (0001) film grown by the same procedure as described above. We conducted the XPS experiment at Centro do Desenvolvimento da Tecnologia Nuclear. The measurements were carried out in an ultrahigh vacuum chamber (base pressure is better than $2.0 \times 10^{-10}$ mbar) using an Al K$\alpha$ x-ray source with the output power set at 300 W and a VG Microtech hemispherical electron energy analyzer CLAM 2/1 VU. Normal emission scans with 50 eV pass energy were acquired. Following the Co and Ru XPS signal before and after the annealing procedure, we did not observe any evidence of Co interdiffusion (see Supplementary Fig. S.4).

**Real-space imaging.** In the SPLEEM system, real-space images with magnetic contrast along three orthogonal directions are acquired, corresponding to the out-of-plane magnetization direction and two orthogonal in-plane axes, as shown in Supplementary Fig. S.5a-c. The lateral spatial resolution of the SPLEEM at Berkeley lab is ~15 nm, while the measured DW width in the systems studied here is between 150 nm to 350 nm. The energy of the incident electron beam was set to 3.6 eV for Ru/Co/graphene and 5 eV for Ru/[Co/Ni]$_n$; these values were chosen to optimize the magnetic contrast. The images are represented in grey scale, where a black and white contrast correspond to the magnetization vector pointing into the film plane (+$M_z$) and out of the plane (-$M_z$), respectively, as shown in Supplementary Fig. S.5a. From the in-plane images, Figs. S.5b and 5c, the DW structure can be seen as narrow lines along the magnetic DWs. To highlight DW spin structures, the triplets of SPLEEM images representing out-of-plane and orthogonal in-plane magnetization components are combined into single compound images, as shown in Supplementary Figure S.5e. In this projection, colours represent the in-plane magnetization as indicated by the colour wheel (inset), black and grey values represent the perpendicular magnetization component, $+M_z$ and $-M_z$, respectively.

**Analysis of chirality.** The method to analyze DW chirality from the SPLEEM images is the same as described by G. Chen, et al.[45]. First, along all DWs the DW normal direction **n** is determined from the out-of-plane magnetization SPLEEM images, where **n** is defined as a vector pointing from spin-down ($-M_z$) to spin-up ($+M_z$) domains. Then, at all pixels along the DW centerlines, the in-plane magnetization direction, (**m**), is measured from the grey values of the two in-plane SPLEEM images. To improve the signal-noisy ratio, in this step each pixel is averaged with its three nearest neighbour pixels. Finally, we compute the angle α, defined as the angle between **m** and **n** (inset of Fig. 3c), and we calculate its distribution along all DW centerlines, as represented by the histograms.

**ACKNOWLEDGEMENTS**

This work was supported by the European Union's Horizon 2020 research and innovation programme under grant agreement No. 696656 (GRAPHENE Flagship), the ANR ULTRASKY, SOSPIN. *Ab initio* calculations used resource of GENCI-CINES with grant No. C2016097605. Work at the Molecular Foundry was supported by the Office of Science, Office of Basic Energy Sciences, of the U.S. Department of Energy under Contract No. DE-AC02-05CH11231. Alexandre A. C. Cotta, Waldemar A. A. Macedo and Edmar A. Soares acknowledge the support of Brazilian agencies CAPES, CNPq and FAPEMIG. We thank V. Cros and A. Thiaville for fruitful discussions and comments.

**AUTHOR CONTRIBUTIONS**

H.X.Y. and G.C. conceived the study. H.X.Y and S.A.N. performed the *ab-initio* calculations with help of M.C. H.X.Y., M.C., S.A.N. and A.F. analyzed and interpreted the *ab-initio* results. G.C. and A.A.C.C. carried out the SPLEEM measurements. A.K.S. supervised the SPLEEM facility. G.C., A.A.C.C., A.T.N. analyzed the SPLEEM results. G.C. derived DMI strength from experimental data. G.C., A.A.C.C., A.T.N., A.K.S., E.A.S, W.A.A.M., interpreted and discussed the experiment result. A.A.C.C., E.A.S, W.A.A.M. performed XPS measurement. H.X.Y and G.C. prepared the manuscript with help from A.A.C.C., A.K.S., S.A.N. and M.C. All authors commented on the manuscript.

**COMPETING INTERESTS STATEMENT**

The authors declare that they have no competing financial interests.

**Figure Captions**

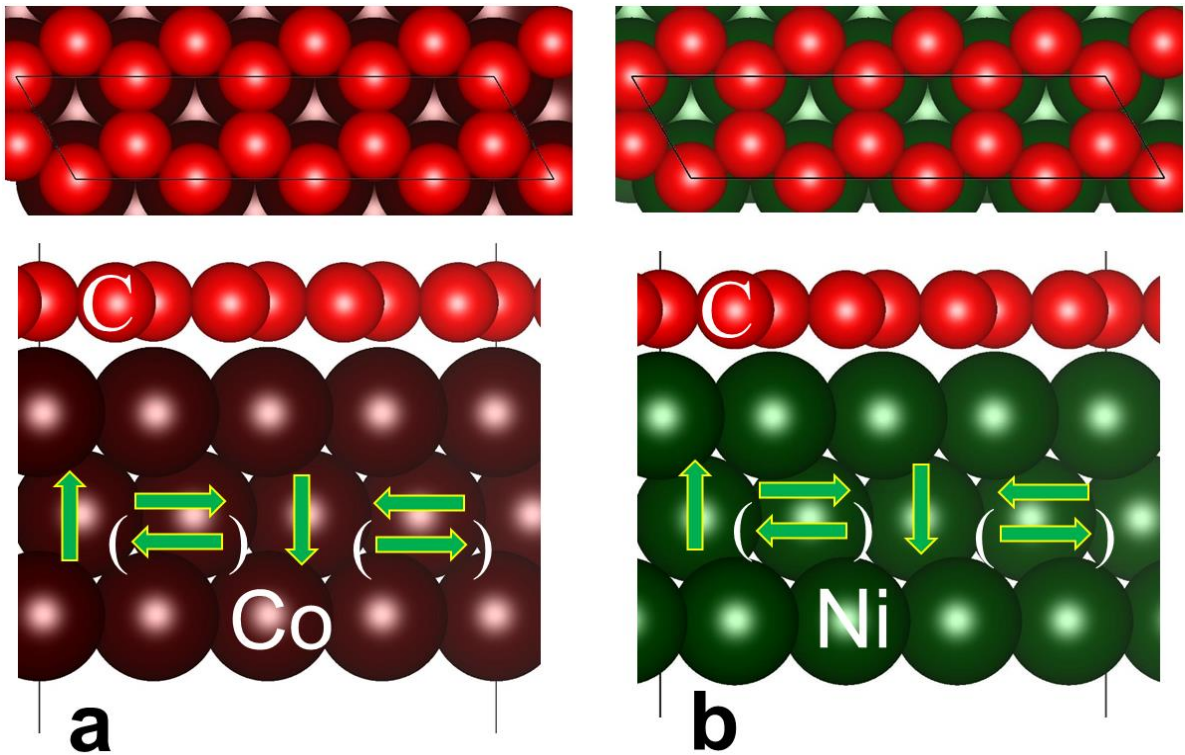

**Figure 1 Crystal and spin configurations of graphene coated Co and Ni films used for DMI calculations. a,** Top- and side-view of graphene on hcp Co(0001) and **b**, top- and side-view of graphene on fcc Ni(111) surface. Red, purple and green balls represent carbon, cobalt and nickel atoms, respectively. Clockwise (anticlockwise) spin configurations are schematically shown by arrows.

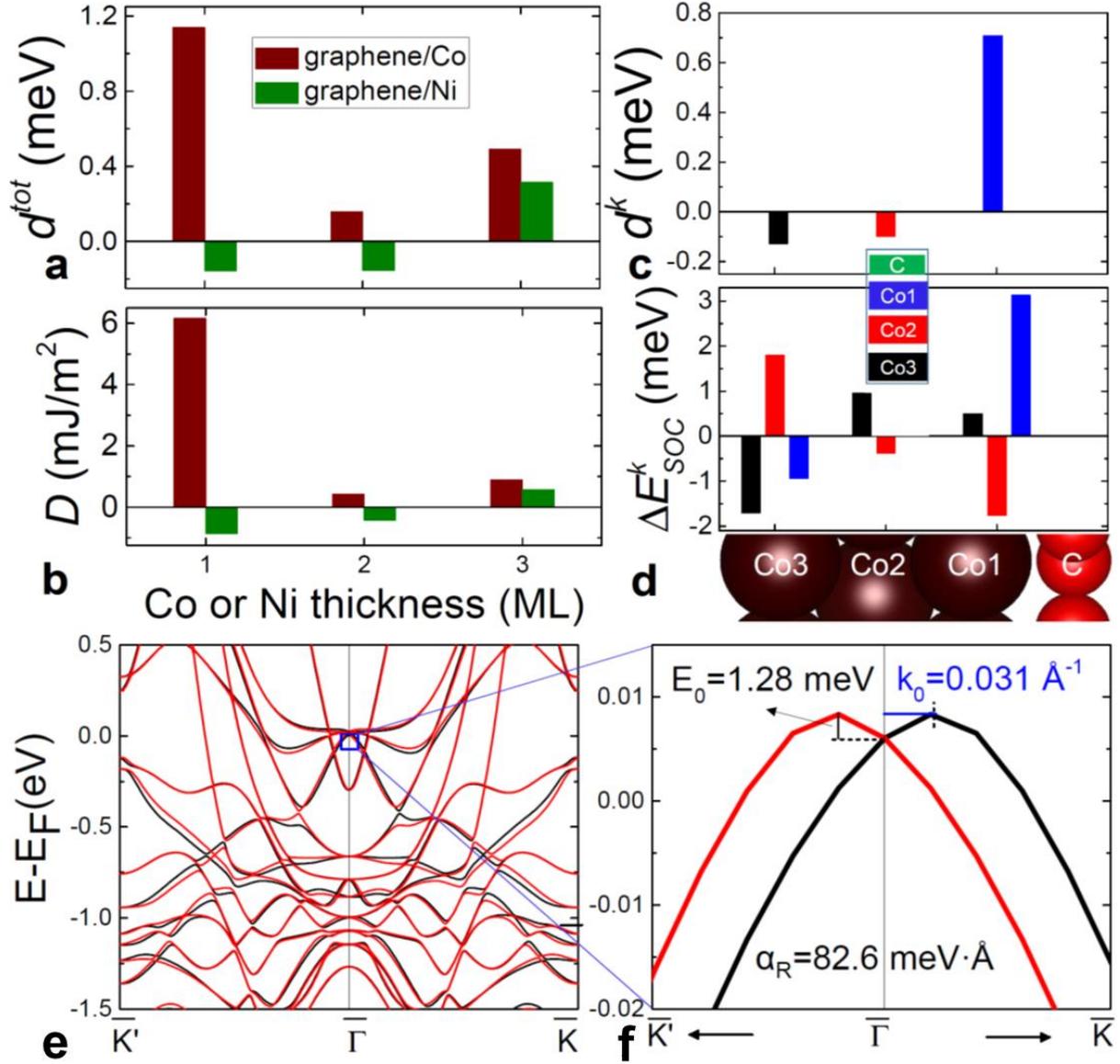

**Figure 2 Anatomy of DMI for graphene/Co and graphene/Ni bilayers. a,** Total DMI coefficient $d^{tot}$ and **b,** micromagnetic DMI coefficient $D$, as a function of FM film thickness for graphene/Co (brown bars) and graphene/Ni (green bars) slabs. **c,** Layer-resolved DMI coefficient $d^k$ of the $k^{th}$ layer for graphene/Co(3ML) slab. **d,** Atomic layer resolved localization of the associated spin-orbit energy $\Delta E_{SOC}^k$. As it is seen, the large DMI coefficient of the Co1 layer (blue bar in **c**) is associated with large variations of the SO energy $\Delta E_{SOC}^{Co1}$ in the Co1 layer (see the corresponding blue bar in **d**). **e** and **f**, Band structures for graphene/Co(3ML) slab with the magnetization axis along <11$\bar{2}$0> (black) and <$\bar{1}\bar{1}$20> (red) used to estimate the Rashba splitting. The corresponding DMI is found to be about 0.30 meV.

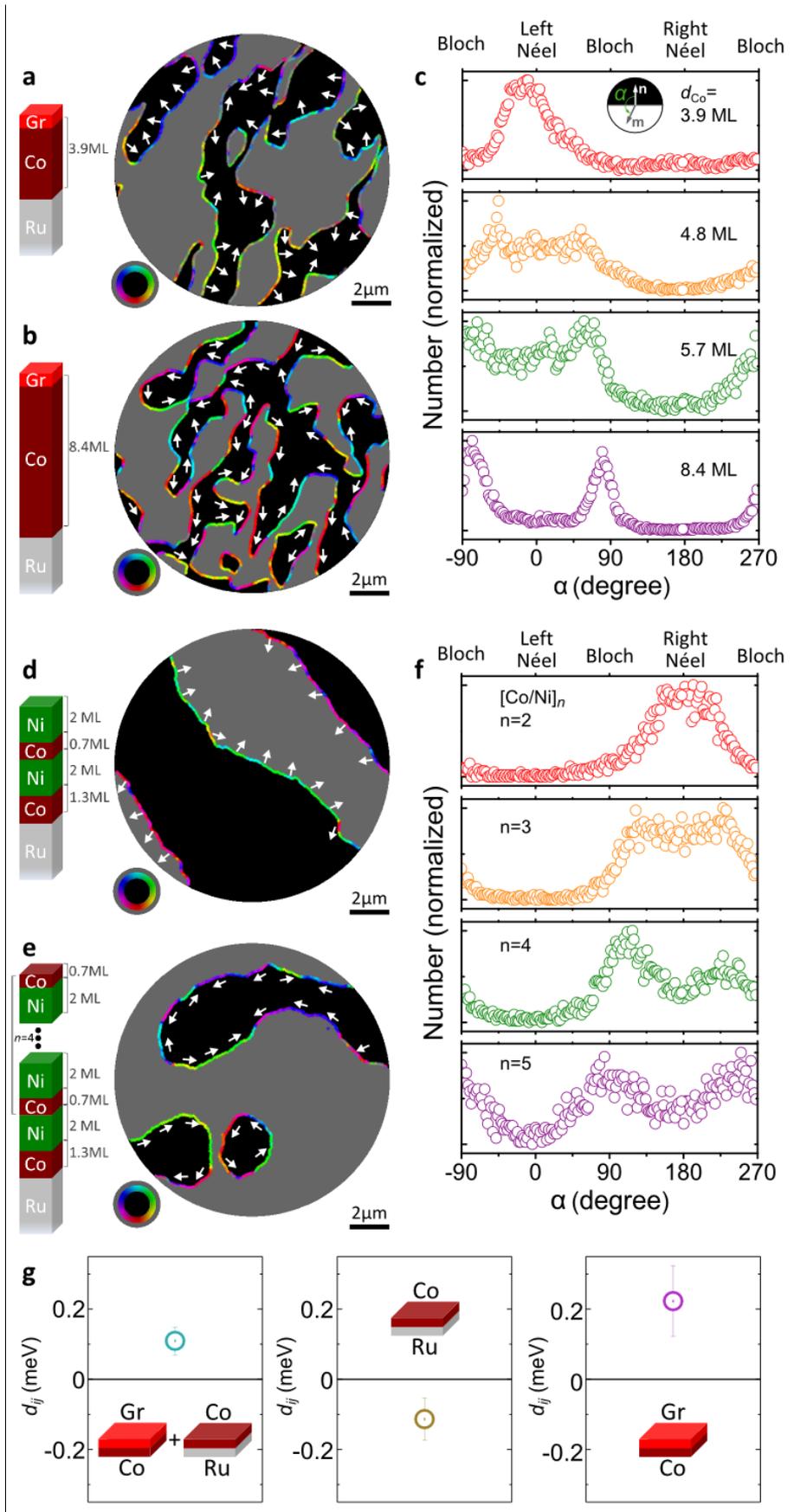

**Figure 3 Experimental measurement of DMI in graphene/Co by using SPLEEM. a,b,** Compound SPLEEM images of graphene/Co/Ru. Scale bar, 2μm. **c,** Co thickness dependent histogram of the angle *α* counted pixel-by-pixel at the DW boundary in graphene/Co/Ru(0001) shows the evolution of the chirality from a left-handed Néel wall (single peak at 0°) to an achiral Bloch wall (double peaks at ±90°). Inset shows the definition of the angle *α*, where **m** is the in-plane direction of the DW magnetization, and **n** is the in-plane vector normal to the domain boundary and always points from grey domains to black domains. **d,e,** Compound SPLEEM images of [Ni/Co]$_n$/Ru(0001). Scale bar, 2μm. **f,** Co thickness dependent histogram of the angle *α* in [Ni/Co]$_n$/Ru(0001) shows the evolution of the chirality from a right-handed Néel wall to an achiral Bloch wall. **g,** Calculated DMI vector ***d***$_{ij}$.

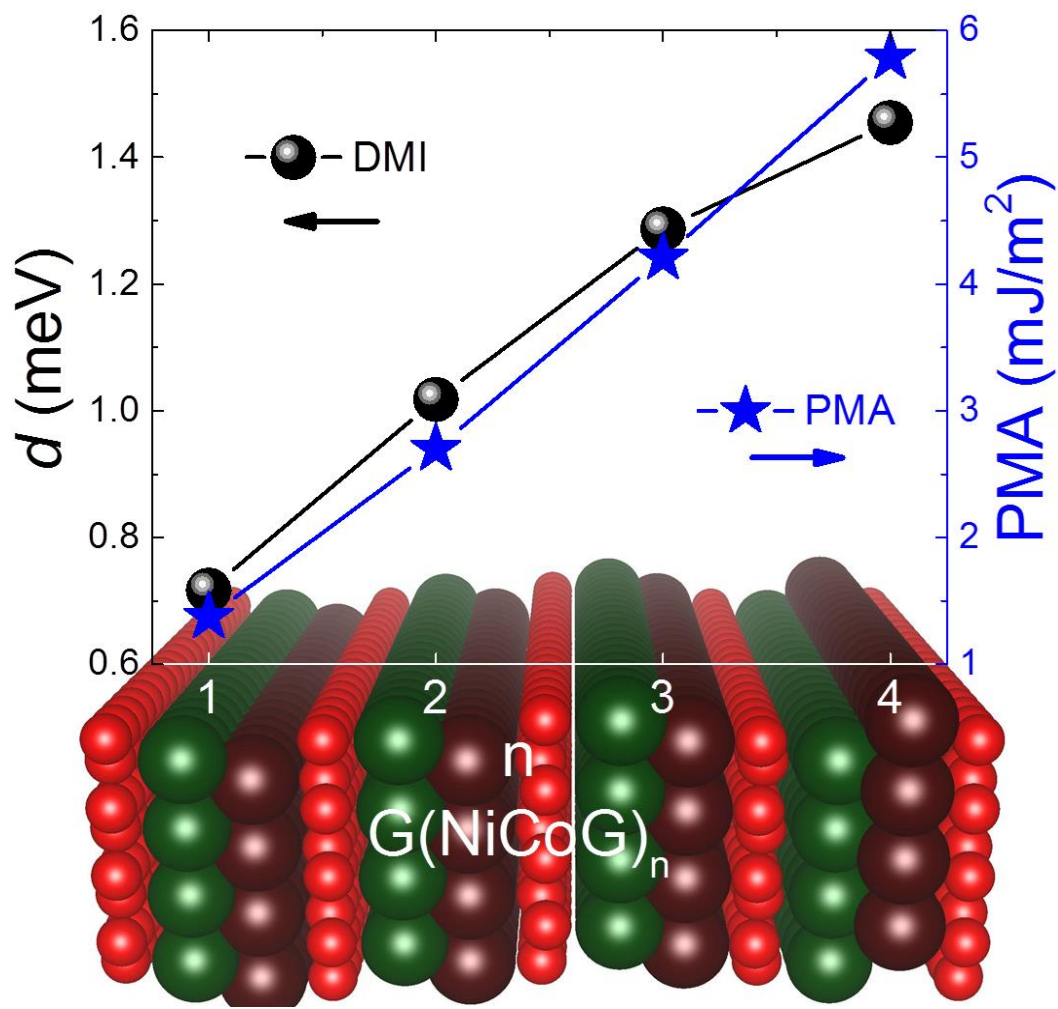

**Figure 4** DMI and PMA for the multilayer of graphene/NiCo/graphene as a function of the junction number *n*.